%% file: tdm-graph.tex
\documentclass[sigconf]{acmart}

\AtBeginDocument{%
  \providecommand\BibTeX{{%
    \normalfont B\kern-0.5em{\scshape i\kern-0.25em b}\kern-0.8em\TeX}}}

\copyrightyear{2021} 
\acmYear{2021} 
\setcopyright{acmcopyright}
\acmConference{DLP-KDD 2021}{August 15, 2021}{Singapore}

\usepackage{graphicx}
\usepackage{multirow,makecell,url,footmisc}
\usepackage{enumitem}
\usepackage{lineno}
\usepackage{textcomp}
\usepackage{booktabs} 
\usepackage{arydshln}
\usepackage{url}
\usepackage{float}
\usepackage{algorithmic}
\usepackage{algorithm}
\usepackage{boldline}
\usepackage{stfloats}
\usepackage[keeplastbox]{flushend}

\makeatletter
\def\adl@drawiv#1#2#3{%
           \hskip.5\tabcolsep
           \xleaders#3{#2.5\@tempdimb #1{1}#2.5\@tempdimb}%
                   #2\z@ plus1fil minus1fil\relax
           \hskip.5\tabcolsep}
\newcommand{\cdashlinelr}[1]{%
     \noalign{\vskip\aboverulesep
              \global\let\@dashdrawstore\adl@draw
              \global\let\adl@draw\adl@drawiv}
     \cdashline{#1}
     \noalign{\global\let\adl@draw\@dashdrawstore
              \vskip\belowrulesep}}
\makeatother

\begin{document}

\title{Context-aware Tree-based Deep Model for Recommender Systems}

\author{Daqing Chang$^*$}
\affiliation{%
  \institution{Alibaba Group}
}
\email{daqing.cdq@alibaba-inc.com}
\author{Jintao Liu$^*$}
\affiliation{%
  \institution{Tsinghua University}
}
\email{liujt17@mails.tsinghua.edu.cn}
\author{Ziru Xu}
\affiliation{%
  \institution{Alibaba Group}
}
\email{ziru.xzr@alibaba-inc.com}
\authornote{Daqing Chang, Jintao Liu, Ziru Xu contributed equally to this work.}
\author{Han Li}
\affiliation{%
  \institution{Alibaba Group}
}
\email{lihan.lh@alibaba-inc.com}
\author{Han Zhu}
\affiliation{%
  \institution{Alibaba Group}
}
\email{zhuhan.zh@alibaba-inc.com}
\author{Xiaoqiang Zhu}
\affiliation{%
  \institution{Alibaba Group}
}
\email{xiaoqiang.zxq@alibaba-inc.com}
\renewcommand{\shortauthors}{Chang, Liu, Xu et al.}

\begin{abstract}
    
    How to predict precise user preference and how to make efficient retrieval from a big corpus are two major challenges of large-scale industrial recommender systems.
    In tree-based methods, a tree structure $\mathcal{T}$ is adopted as index and each item in corpus is attached to a leaf node on $\mathcal{T}$. Then the recommendation problem is converted into a hierarchical retrieval problem solved by a beam search process efficiently. 

    In this paper, we argue that the tree index used to support efficient retrieval in tree-based methods also has rich hierarchical information about the corpus. 
    Furthermore, we propose a novel  context-aware tree-based deep model (ConTDM) for recommender systems.
     In ConTDM, a context-aware user preference prediction model $\mathcal{M}$ is designed to utilize both horizontal and vertical contexts on $\mathcal{T}$.
     Horizontally, a graph convolutional layer is used to enrich the representation of both users and nodes on $\mathcal{T}$ with their neighbours.
     Vertically, a parent fusion layer is designed in $\mathcal{M}$ to transmit the user preference representation in higher levels of $\mathcal{T}$ to the current level,
     grasping the essence that tree-based methods are generating the candidate set from coarse to detail during the beam search retrieval. 
     Besides, we argue that the proposed user preference model in  ConTDM can be conveniently extended to other tree-based methods for recommender systems. 
    Both experiments on large scale real-world datasets and online A/B test in large scale industrial applications show the significant improvements brought by ConTDM.
    
\end{abstract}



\maketitle

\input{section1} 
\input{section2} 
\input{section3} 
\input{section3_1}
\input{section5} 
\input{section6} 

\bibliographystyle{ACM-Reference-Format}
\bibliography{tdm-graph}

\newpage
\end{document}

%% file: section1.tex
\section{Introduction}\label{section:Introduction}

 Recommendation problem is generally to retrieve for a candidate set comprised by users' most preferred items from the entire corpus.
 In large-scale industrial recommender systems, making precise user preference prediction and efficient retrieval from a big corpus is extremely important.
 The linear retrieval complexity of traversing the entire corpus is usually unacceptable. 
 Together with a user preference prediction model, a proper index structure is usually necessary in retrieving the candidate set. 

 Recently, vector representation learning methods have become more and more popular in recommender systems \cite{covington2016deep,wang2019neural,lv2019sdm,koren2009matrix}. In these methods, both users and items are firstly represented by vectors in a same space. Then the inner-product of the user vector and the item vector is used as the metric of user-item preference. 
 As a main benefit, the candidate generation for these methods equals to a classic k-nearest neighbour problem, which can be accelerated by quantization-based index \cite{johnson2019billion,liu2005investigation}, hierarchical graph index\cite{malkov2018efficient} etc..
 Many efforts such as sequential model \cite{lv2019sdm}, graph convolutional network \cite{wang2019neural} have been made to learn better user vectors and item vectors. 
 However, the inner-product form of user-item preference modeling required by these methods is still a bottleneck for improving user preference prediction accuracy\cite{zhu2018learning,he2017neural}. 

\begin{figure}[!htbp]
    \centering
    \includegraphics[width=0.45\textwidth]{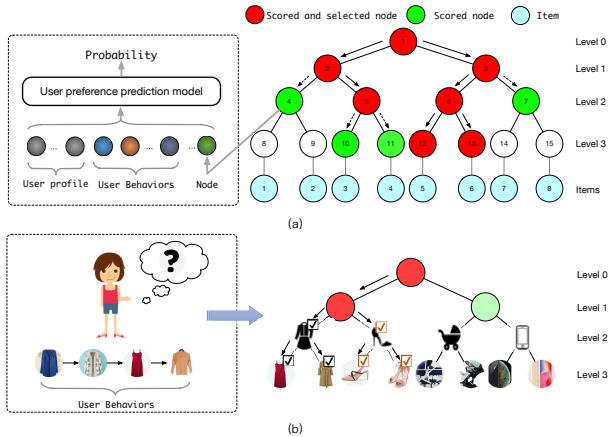}
    \caption{(a). A general framework of tree-based model for recommender systems. In retrieval, a top-down beam search on the tree index is firstly processed to generate the final candidate set guided by a user preference prediction model. Then items indexed on nodes in the beam of the leaf level are generated as candidate sets. In this example, the beam size is 2 and item 5, 6 are finally selected for recommendation. (b) A toy example to show the value of contexts between nodes on $\mathcal{T}$.}\label{fig:example}
\end{figure}

 To break this bottleneck, a tree structure is used as index in tree-based methods \cite{morin2005hierarchical,zhu2018learning,zhu2019joint,zhuo2020learning}. 
 The general framework of tree-based methods for recommender systems is shown in Figure \ref{fig:example}(a). 
 Firstly, each item in the corpus is carefully indexed to a leaf node of $\mathcal{T}$  and a user node preference prediction model $\mathcal{M}$ is trained. 
 $\mathcal{T}$ is usually a full binaray tree and the relationship between leaf nodes and items in corpus can be made by clustering \cite{zhu2018learning} or joint learning \cite{zhu2019joint}.
 In retrieval, a top-down beam search process on $\mathcal{T}$ guided by $\mathcal{M}$ is used in candidate generation, which complexity is logarithmic w.r.t. the corpus size.
 With tree index, restrictions on the structure of $\mathcal{M}$ required by the kNN-based retrieval are removed and many advanced user preference prediction model such as Deep Interest Network \cite{zhou2018deep}, Wide \& Deep \cite{cheng2016wide}, Deep Interest Evolution Network \cite{zhou2019deep}, xDeepFM \cite{lian2018xdeepfm} can be naturally used to achieve better user preference accuracy \cite{zhu2018learning}.

 However, all models above are originally designed for general user preference prediction and the useful contexts between nodes on tree index are not fully considered.
 Since nodes with a common parent are equally represented by this parent both in training and retrieval, each node on $\mathcal{T}$ is actually an abstraction of its children.
 The beam search retrieval is actually to generate the final candidate set from coarse to detail.
 Intuitively, contexts between nodes on $\mathcal{T}$ should be useful auxiliary knowledge in user preference prediction.
 To better illustrate this, a toy example is shown in Fig \ref{fig:example}(b). 
 According to the girl's historical behaviors, she probably prefers cloths at this time. 
 Suppose $\mathcal{M}$ has rightly predicted  the left node in level 2 as the girl's coarse preference, this vertical context should be useful for $\mathcal{M}$ to generate the final candidate set in leaf level.
 Besides, relationships between nodes on the same level are also useful horizontal contexts in prediction. 
 For this toy example, shoes are also probably preferred by the girl in Fig \ref{fig:example} because clothes and shoes are usually correlative things in daily life.
 However, shoes may not be chosen by $\mathcal{M}$ if only historical behaviors are used as features in prediction.
 

 In this paper, we propose a Context-aware Tree-based Deep Model (ConTDM) to utilize both vertical and horizontal contexts on tree index in user preference prediction for tree-based recommender systems. Generally, a novel context-aware user node preference prediction model $\mathcal{M}$ is proposed in ConTDM. Our main contributions are listed as follows: 
 
\begin{itemize}[topsep=0pt,parsep=0pt,partopsep=0pt,itemsep=0.25em,leftmargin=1em]
    \item Horizontally, contexts between nodes on the same level of $\mathcal{T}$ are aggregated with a graph convolutional layer. For tree-based models, a hierarchical graph structure is necessary to utilize vertical contexts on all levels of $\mathcal{T}$. 
    We propose a novel  hierarchical graph construction algorithm according to raw user behavior sequences and tree index.
    \item Vertically, a parent fusion layer is designed in $\mathcal{M}$. In ConTDM, we take the user preference representations predicted on higher levels of $\mathcal{T}$ as vertical contexts. 
        Through the parent fusion layer, they are imported as an auxiliary input in prediction on the current level.
    \item We argue that the proposed user preference model can be conveniently extended to other tree-based methods. The training of ConTDM in this paper follows the framework proposed in TDM \cite{zhu2018learning} without loss of generality. Offline experiments and ablation study on open data sets shows the significant improvements of ConTDM compared with baseline methods. 
    \item ConTDM has been applied in full production to  the display advertising scenario of Guess What You Like column of Taobao App Homepage at the candidate generation stage. Online A/B test shows the significant improvements on click-through rate (CTR) and revenue per mille (RPM), which are key performance indicators for online display advertising.
\end{itemize}

 The rest of the paper is organized as follows: We introduce related works in Section 2. The proposed context-aware user preference prediction model and the training framework of ConTDM are introduced in Section 3. Experimental results are analysed in Section 4. We conclude our work in Section 5. 
 

%% file: section2.tex
\section{Related Work}\label{section:Relatedwork}

 Vector representation learning methods beginning from matrix factorization based collaborative filtering have been widely used in recommender systems \cite{koren2009matrix,wang2015collaborative,xue2017deep,covington2016deep,koren2008factorization,wang2019neural}. In these methods, both users and items are mapped to vectors in the same space and 
the user-item preference is measured by the inner-product of user and item vectors.
 As an early representative work, a multi-layer fully connected network is used to project users and items into a latent space in industrial YouTube video recommender systems \cite{covington2016deep}. 
 As a main benefit, the candidate retrieval for these methods equals to a k-nearest neighbour problem, which can be accelerated by quantization based index \cite{johnson2019billion,liu2005investigation}, hierarchical graph index\cite{malkov2018efficient} etc..
 Variants of improvements have been made in learning the vector mappings. For example, Lv et al. \cite{lv2019sdm} use both recurrent neural network and attention mechanism in learning user vectors. Wang et al. \cite{wang2019neural} use a graph neural network to aggregate local information from the graph structure. 
 However, the simple inner-product form of user preference modeling required by the kNN-based retrieval is still a key bottleneck for recommendation accuracy due to its limited learning capacity \cite{zhu2018learning,he2017neural}. Many other user preference models that has been shown to be effective such as Deep Interest Network \cite{zhou2018deep}, Deep \& Wide \cite{cheng2016wide}, Deep Interest Evolution Network \cite{zhou2019deep}, xDeepFM \cite{lian2018xdeepfm} usually can not be applied directly in these methods.  

\begin{figure*}[t]
    \centering
    \includegraphics[width=0.8\textwidth]{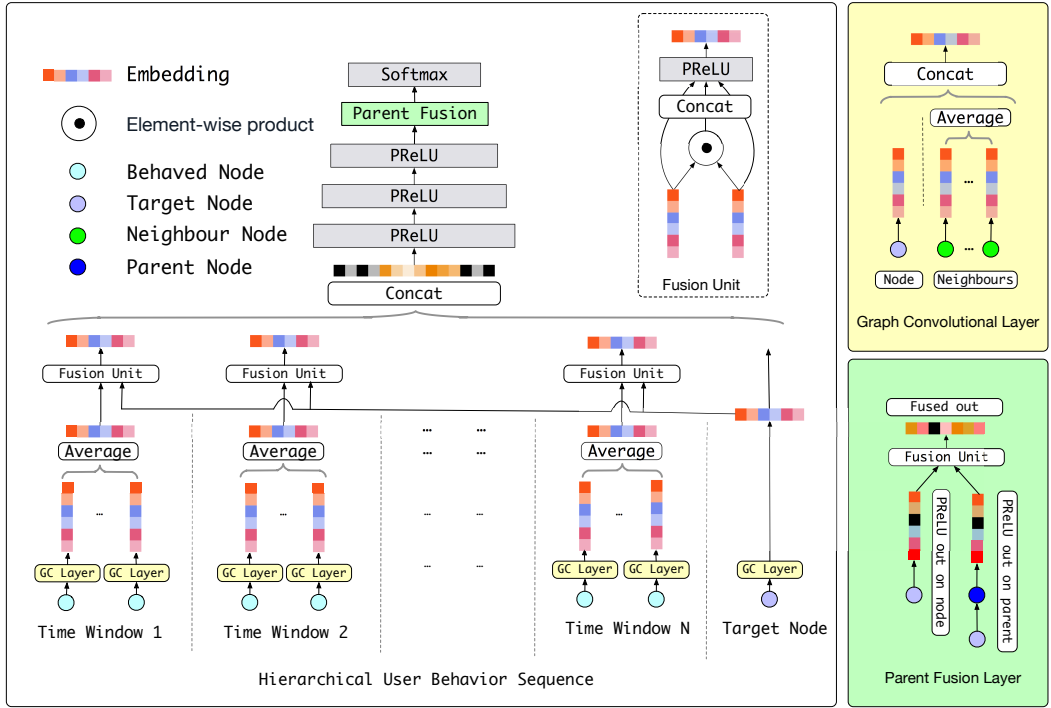}
    \caption{The backbone of user preference prediction model in ConTDM. Highlights: a) A graph convolutional layer is used to aggregate horizontal contexts on the tree index. Both the user embeddings and target node embedding are enhanced by this layer.
     b) The parent fusion layer takes both the output of the multi-layer fully connected network on the target node and its parent as inputs and a fusion unit is used to utilize vertical contexts on the tree index. } \label{fig:ConTdm}
\end{figure*}
 
 In past years, tree-based methods are actively studied in the field of extreme classification \cite{prabhu2014fastxml,prabhu2018parabel,weston2013label,daume2017logarithmic,morin2005hierarchical,tanno2018adaptive,you2019attentionxml}, which is also closely related with recommender systems \cite{jain2016extreme,prabhu2018parabel}.   
 To break the bottleneck of kNN-based methods, tree-based methods \cite{zhu2018learning,zhu2019joint,zhuo2020learning} take a tree structure as index and each item in corpus is attached to a leaf node by clustering \cite{zhu2018learning} or joint learning \cite{zhu2019joint}.
 A beam search process from top to bottom is used in retrieval and a logarithmic complexity w.r.t. the corpus size is achieved.
 In these methods, restrictions on the form of user preference prediction model is removed and the use of arbitrary advanced user preference models is enabled to improve the recommendation accuracy.
 In training,
 a joint optimization framework of the user preference prediction model and the tree index \cite{zhu2019joint} is proposed. 
 More recently, An optimal beam search aware training framework of tree-based deep models \cite{zhuo2020learning} is proposed to eliminate the mismatch in training and retrieval. 

 On the other hand, graph-based methods have also attracted much attention in recommender systems \cite{zhu2019aligraph,ying2018graph,wang2018billion,fan2019graph,zhao2017meta,wu2019session}. The general idea of graph-based methods is to make effective information aggregation from the local subgraph with a well constructed graph structure. Variants of aggregator architectures have been proposed in literatures \cite{hamilton2017inductive,velivckovic2017graph,kipf2016semi}. 
 In industrial community, Ying et al. \cite{ying2018graph} propose the PinSage algorithm used in Pinterest by adding a random walk based neighbour sampling strategy to GraphSage \cite{hamilton2017inductive}.
 By importing side information to DeepWalk \cite{perozzi2014deepwalk}, Wang et al. \cite{wang2018billion} propose the graph embedding algorithm used in Alibaba to generate the item embeddings. They are used to calculate the similarity matrix for subsequent Item-CF \cite{sarwar2001item} based recommendation.     
 However, to our best knowledge, there is no existing work applied in industrial community simultaneously utilizing the superiority of both tree-based methods and graph-based models.

%% file: section3.tex
\section{Context-aware Tree-based Deep Model}\label{section:Method}


 In this section, we introduce the proposed Context-aware Tree-based Deep Model for recommender systems.
 As a main highlight of ConTDM, the proposed context-aware user node preference prediction model is given in Section \ref{sec:model}. We show the training framework of ConTDM in Section \ref{sec:framework}.

\subsection{Context-aware User Preference Prediction Model}\label{sec:model}

 In tree-based methods for recommendation, a tree structure $\mathcal{T}$ is used as index and each item in corpus is carefully indexed to a leaf node on $\mathcal{T}$ by clustering \cite{zhu2018learning}, joint learning \cite{zhu2019joint} etc., 
 as introduced in Section \ref{section:Introduction} and Fig \ref{fig:example}. Then a beam search process guided by a user node preference prediction model $\mathcal{M}$ is made to generate the candidate set for recommendation in retrieval. 
 Obviously, the accuracy of user node preference prediction has great influence on the final recommendation quality for tree-based methods. 
 Inspired by the essence of tree-based methods is to generate the candidate set from coarse to detail, we hope to fully utilize the rich hierarchical information on the tree index about the corpus in designing the structure of $\mathcal{M}$. 
 More specifically, both vertical and horizontal contexts contained on $\mathcal{T}$ are properly utilized to improve the accuracy of $\mathcal{M}$ in ConTDM.
 The backbone of $\mathcal{M}$ is shown in Fig \ref{fig:ConTdm}.

 Denote $(c_1,c_2\cdots c_m)$ as the historical behavior sequence of the user $u$ and denote $b_j(c)$ as the ancestor of node $c$ on level $j$ of $\mathcal{T}$.
 We use hierarchical user behavior sequences \cite{zhu2019joint} that have been shown effective in tree-based methods for recommender systems as our user features.
 More exactly, the user feature is represented as $(b_j(c_1),b_j(c_2)\cdots$ $b_j(c_m))$ when the target node in prediction is on level $j$ of $\mathcal{T}$.
 Other useful features such as user profiles can be conveniently added if needed. 
 In estimating the user node preference probability,
 the embeddings of both the hierarchical user behavior sequence and the target node are firstly put into a graph convolutional layer to import horizontal contexts on $\mathcal{T}$. 
 Then user historical behaviors are divided into different time windows and  
 graph embeddings in the same time window are averaged to reduce the network complexity, which is optional to meet the practical time constraint in large scale industrial application without much loss of the effect at the same time.
 Next, a fusion unit is used to fuse the graph embeddings in user behaviors with the graph embedding of the target node similarly as the attention mechanism, which can also been employed in practice.
 We use the fusion unit here for consistency. 
 After this,
 the user embeddings and the target embedding are concatenated as the input of a  multi-layer fully connected network with PReLU activation function.
 The outputs of the network on both the target node and its parent are further fused by a parent fusion layer to import vertical contexts.
 Finally, a softmax layer is used to compute the user-node preference probability.
 
 As key components to utilize contexts between nodes, the mechanism and effect of  the graph convolutional layer and the parent fusion layer are further analysed in the following two subsections. 

%% file: section3_1.tex
\begin{figure*}[htbp]
    \centering
    \includegraphics[width=0.95\textwidth]{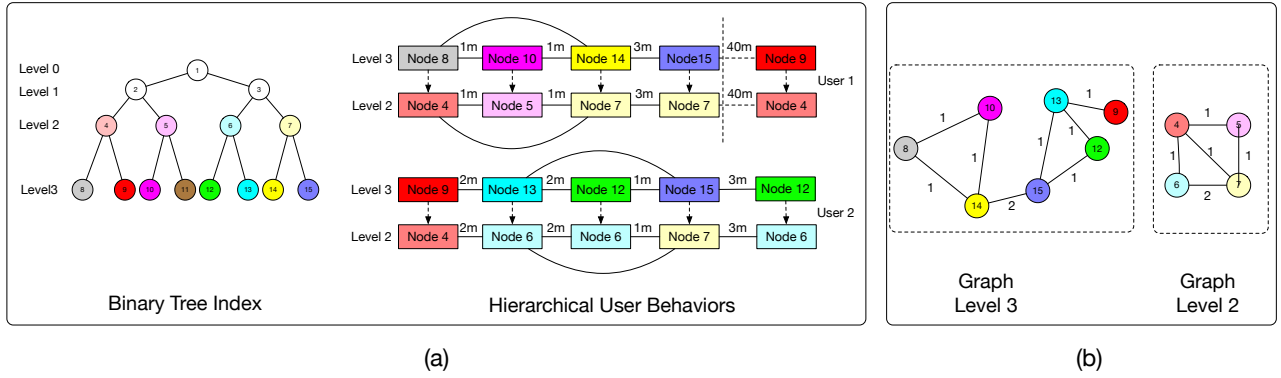}
    \caption{(a) An example to generate edges based on tree index and raw user behaviors. Firstly, hierarchical behavior sequences are contructed. Then 
    an edge is put into the graph between an co-occurrence pair of nodes if the gap of their time stamps is not bigger than a threshold. 
    (b) The constructed hierarchical graph based on sequences in (a). The weight of an edge is the count of co-occurrence between its two endpoints. 
    }\label{fig:Graph}
\end{figure*}

\subsubsection{\textbf{Graph Convolutional Layer}}\label{sec:gclayer}

 Horizontally, the relevance between nodes on the same level of $\mathcal{T}$ can be used to enrich both the user and the target node features.
 As discussed in Section \ref{section:Relatedwork}, graph structures have been proven to be powerful to grep the relationship between items in corpus.
 
 In ConTDM, a non-parameterized graph convolutional layer based on GraphSage\cite{hamilton2017inductive} is used considering the efficiency in training and inference,
 as shown in Fig \ref{fig:ConTdm}.
 For each node $n$ on the tree index, 
 its \textbf{g}raph \textbf{e}mbedding is computed by:
 \begin{equation}
     Emb_{ge}(n) = Concat(Emb(n), Avg(Emb(Neigh(n))) 
 \end{equation}
 where $Emb(n)$ is the embedding of node $n$ and  $Neigh(n)$ is the set of $n's$ neighbours on the graph. 
 The graph convolutional layer concatenates the averaged embeddings of $n's$ neighbours with $n's$ embedding. 
 The purpose is to import the neighbour context as well as highlight the role of $n$.
 
 With the graph convolutional layer, the representation of each node is enriched by its neighbours.
 On the one hand, the impact of sparsity in training data that has been widely observed in recommender systems \cite{sarwar2001item} is alleviated.
 On the other hand, the scope of the user feature could jump out of historical behaviors and the diversity of the candidate set is promoted, which is usually preferred by most recommender systems. 
 Besides, it is worth to point out that 
 graph embeddings of all nodes can be computed efficiently before the trained model is put online according to the backbone of $\mathcal{M}$.
 Therefore, as another benefit brought by the graph convolutional layer, the time cost of online forward computation which is usually strictly bounded in industrial recommendation scenario can be saved.

 \begin{algorithm}[ht]
\caption{Building Hierarchical Graph in ConTDM}\label{alg:BuildGraph}
\begin{algorithmic}[2]
    \REQUIRE Tree index $\mathcal{T}$ with max level $l_{max}$, raw user behavior sequences of training data $\{(c_1,c_2\cdots c_m)\}$, max number of neighbours $k$, max time interval $t$
    \STATE $\mathcal{G} \gets \emptyset$
    \FOR {$l=l_{max}, \dots, 1$}
        \FOR {each raw sequence $(c_1,c_2 \cdots c_m)$}
			\STATE $(b_l(c_1),b_l(c_2)\cdots b_l(c_m)) \gets$ Trace each item in raw sequence up to level $l$ of $\mathcal{T}$.
			\IF{node $n_i$ and $n_j$ are both contained in $(b_l(c_1),b_l(c_2)\cdots b_l(c_m))$ \textbf{and} the time interval between them $\leq t$ minutes}
                \STATE Increase the weight of undirected edge $(n_i, n_j)$ in $\mathcal{G}$ by 1.
		    \ENDIF
		\ENDFOR
        \FOR {each node $n$ in level $l$ of $\mathcal{T}$}
            \STATE $\{n_i\}_{i=1}^K \gets$ Pick the other endpoint of all undirected edges containing $n$ in $\mathcal{G}$.
			\STATE Sort  $\{n\}_{i=1}^K$ in the descending order of $(n, n_i)$'s weight.
			\FOR {$i=1, \dots, \min(k, K)$}
				\IF{edge $(n, n_i)$ not in $\mathcal{G}$}
					\STATE Add undirected edge $(n, n_i)$ to $\mathcal{G}$.
				\ENDIF
			\ENDFOR
		\ENDFOR
	\ENDFOR
\ENSURE Undirected hierarchical graph $\mathcal{G}$.
\end{algorithmic}
\end{algorithm}
 
 Obviously, the quality of the graph matters. Besides, a hierarchical graph is required in ConTDM to aggregate horizontal contexts on all levels of $\mathcal{T}$.
 We build the hierarchical graph according to the co-occurrence of nodes in hierarchical user behavior sequences $\{(b_j(c_1),b_j(c_2)\cdots$ $b_j(c_m))\}$.
 The  process is explained in Fig \ref{fig:Graph}(a).
 The user behavior sequence is firstly divided into sessions according to the gap between timestamps.
 Then an edge between each co-occurrence pair in the same session is added to the graph. The weight of an edge is the count of all co-occurrence pairs between its two endpoints.
 Generally, our main idea is that two items behaved by a user sequentially are probably related with each other 
 and the hierarchical behavior sequence is the abstract user behaviors on different levels of $\mathcal{T}$.
 A formal statement  about the construction of the hierarchical graph is given in Alg \ref{alg:BuildGraph} where $t$ is the threshhold (e.g. $40min$) to divide the sessions in behavior sequences and $k$ is the max number of neighbours used in truncation (e.g. $10$ in practice) to avoid the number of neighbours for hot nodes is too big and only keep the solid relationship as well.

\subsubsection{\textbf{Parent Fusion Layer}} \label{sec:mtl-layer}

 Given a user $u$ and a target node $n_l$ on level $l$ of $\mathcal{T}$, we denote $pa(n_l)$ as the parent node of $n_l$ in level $l-1$
 and denote $v(n_l)$ as the output of the multi-layer fully connected network.
 The parent fusion layer takes $v(n_l)$ and $v(pa(n_l))$ as input and returns the fused user node preference representation through a fusion unit. The detailed formulation is 
 \begin{equation}\label{eq:fuseunit}
         \quad v_{f} = 
         PReLU\left(W* Concat\left(v(n_l),v(n_l)\odot v(pa(n_l)),v(pa(n_l))\right)+b\right).
 \end{equation}
 where $W$ is the weight term and $b$ is the bias term to be optimized.
 With the parent fusion layer, the vertical contexts on higher levels of $\mathcal{T}$ are imported to the current prediction. The impact of the parent fusion layer is shown as follows:

\begin{itemize}
    \item \textbf{Explainability}. In the tree index, each node is an abstraction of its children.
        Traditionally, $v(n_l)$ is used as the input vector of the final softmax layer to compute the final user preference probability,
        which indicates that $v(n_l)$ should contain useful user preference information.
        Besides, the essence of the hierarchical retrieval is to generate the final candidate set from coarse to fine.
        Therefore, the user preference representation on the parent node $v(pa(n_l))$ is a useful coarse-grained auxiliary feature in the prediction of $n_l$.   
    \item \textbf{Non-sparsity}.
        In ConTDM, the training samples for each node in higher levels is much more enriched than lower levels, since the total number of nodes decreases exponentially with the going up of levels on $\mathcal{T}$. By importing contexts in higher levels, the impact of sparsity in lower levels is also largely reduced. 
        \item \textbf{Efficiency}. Since the hierarchical retrieval is made from top to bottom on $\mathcal{T}$, $v(pa(n_l))$ can be efficiently reused in predicting $n_l$. Therefore, the total increase of computation in retrieval is brought by the fusion unit, which is usually acceptable.
\end{itemize}

\subsection{Training Framework}\label{sec:framework}

 With the proposed context-aware user preference prediction model, the training framework of ConTDM is shown in Alg \ref{alg:ConTdm} following the tree-based deep model proposed in \cite{zhu2018learning}. 

\begin{algorithm}[t]
\caption{The Training Framework of ConTDM}\label{alg:ConTdm}
\begin{algorithmic}[1]
    \REQUIRE Initial context-aware user preference model $\mathcal{M}$, tree index $\mathcal{T}$, raw training data $\{(u_i,q_i)\}_{i=1}^N$.
    \STATE Construct the hierarchical graph including all items in corpus and nodes on $\mathcal{T}$ as vertex with Alg \ref{alg:BuildGraph}.
    \FOR{t=1,2$\cdots$ T}
    \STATE Construct training samples used in current iteration.
    \STATE Optimize $\mathcal{M}$ using algorithms such as ADAM \cite{kingma2014adam}.
    \ENDFOR
\ENSURE Trained model $\mathcal{M}$ used for beam search retrieval.
\end{algorithmic}
\end{algorithm}

 The input of Alg \ref{alg:ConTdm} contains an initial context-aware user preference model $\mathcal{M}$, the tree index $\mathcal{T}$ and the raw training data set  $\{(u_i,q_i)\}_{i=1}^N$ where $u_i$ denotes user $i$ and $q_i$ denotes the label item $u_i$ prefers (e.g. $u_i$ clicks $q_i$ before).
 The initial tree index can be constructed by clustering following \cite{zhu2018learning} without loss of generality.
 Before training $\mathcal{M}$, we firstly construct the hierarchical graph $\mathcal{G}$ used by Alg \ref{alg:BuildGraph}. 
 Next, $\mathcal{M}$ is optimized under the total empirical loss as follows \cite{zhu2018learning}:
\begin{equation}\label{loss}
    \mathcal{L}(\boldsymbol{\theta}) = -\sum_{i=1}^N \sum_{j=0}^{l_{max}} \log \hat p \left( b_j (q_i) \vert u_i;\boldsymbol{\theta} \right)
\end{equation}
 where $l_{max}$ is the max level of tree index $\mathcal{T}$. $\boldsymbol{\theta}$ is the parameter of $\mathcal{M}$ to be optimized. $b_j(q)$ returns item $q's$ ancestor node on level $j$ of $\mathcal{T}$.  $\hat p (q\vert u)$ is the estimated probability $u$ prefers $q$ by $\mathcal{M}$. 
 In Eq (\ref{loss}), the total negative logarithm of the estimated user-node probability between each pair $(u_i,q_i)$ and their ancestors is minimized.
 In each iteration, we randomly sample a mini-batch samples from the raw data set and tracing them up to all levels of $\mathcal{T}$ as positive data. 
 Besides, negative sampling \cite{jean2014using,covington2016deep} is used in estimating $\hat p(q\vert u)$ with negative data sampled from the corresponding levels of $\mathcal{T}$.
 
As the proposed context-aware user node preference prediction model in ConTDM only relies on the tree index and raw training behavior sequences, it is convenient to be applied to other existing tree-based frameworks for recommender systems such as \cite{zhu2019joint,zhuo2020learning,morin2005hierarchical}.

%% file: section5.tex
\section{Experiments}\label{section:Experiments}

 In this section, we show both online and offline  performance of ConTDM. 
 Firstly, datasets utilized in offline experiments are briefly summarized. Secondly, we compare the overall performance of ConTDM with other baseline recommendation models to show the effectiveness of the context-aware modeling. Thirdly, ablation study is followed up to help comprehend how each part of ConTDM works in detail. At last, we show the performance of ConTDM in Taobao display advertising platform with real online traffic.

Our offline experiments are conducted with two large-scale real-world datasets: 1) user-book review dataset from Amazon\cite{he2016ups, mcauley2015image}; 2) user-item behavior dataset from Taobao called UserBehavior\cite{zhu2018learning}. The details are as follows:
\begin{itemize}
\item \textbf{Amazon Books}: This dataset is composed by product reviews from Amazon. Here we use its largest subset, i.e., \texttt{Books}. The users with less than 10 books reviewed are excluded. Each review record is in the format of \texttt{(user ID, book ID, rating, timestamp)}.
\item \textbf{UserBehavior}: It is a subset of Taobao user behavior data  containing about 1 million randomly sampled users who had behaviors from November 25 to December 03, 2017. Similar to Amazon Books, only users with at least 10 behaviors are kept. Each user-item behavior is corresponding to a record in the form of \texttt{(user ID, item ID, category ID, behavior type, timestamp)}. All behavior types are treated equal in our experiments.
\end{itemize}
Table 1 summarizes the above two datasets after preprocessing.

\begin{table}[htb]
	\caption{Details of the two datasets after preprocessing. One record is a user-item pair that represents user feedback.}
	\label{tab:dataset}
	\centering
	\begin{tabular}{ccc}
		\toprule
							& \textbf{Amazon Books} & \textbf{UserBehavior} \\
		 \midrule
		 \# of users        & 294,739               & 969,529               \\
		 \# of items        & 1,477,922             & 4,162,024             \\
		 \# of categories   & 2,637                 & 9,439                 \\
		 \# of records      & 8,654,619             & 100,020,395           \\
		\bottomrule
	\end{tabular}
\end{table}

\subsection{Experiment Setup}

In our offline experiments,
 Precision, Recall and F-Measure are used as metrics for performance evaluation of different methods as in most related works for candidate generation in recommender systems. For a user \textit{u}, denote $\mathcal{P}_u(|\mathcal{P}_u| = M)$ as the recalled candidate set and $\mathcal{G}_u$ as the ground truth set. The definitions of these metrics are as follows:
\begin{align*}
    &\text{Precision}@M(u) = \frac{|\mathcal{P}_u \cap \mathcal{G}_u|}{|\mathcal{P}_u|}, ~\text{Recall}@M(u) = \frac{|\mathcal{P}_u \cap \mathcal{G}_u|}{|\mathcal{G}_u|}, \\
    &\text{F-Measure}@M(u) = \frac{2 * \text{Precision}@M(u) * \text{Recall}@M(u)}{ \text{Precision}@M(u) + \text{Recall}@M(u)}.
\end{align*}
	The user average of the above three metrics in testing set are used to compare the following methods:
\begin{itemize}
    \item \textbf{Item-CF} \cite{sarwar2001item}, namely the classic item-based collaborative filtering, maintains an item-item matrix measuring similarities between pairs of items. The recommended items are generated according to the user's historical behaviors and the matrix.
    \item \textbf{YouTube product-DNN} \cite{covington2016deep}, the representative work of kNN-based methods, is a practical method used in YouTube video recommendation.
    The inner-product of the learnt user and item's vector representation denotes the preference.
    \item \textbf{HSM} \cite{morin2005hierarchical} is short for the hierarchical softmax model, which utilize multiplication of level-wise conditional probabilities to obtain item preference probability without the normalization term.
    \item \textbf{TDM} \cite{zhu2018learning} is a representative tree-based deep model for recommender systems. The backbone of its user preference prediction model is comprised by an attention layer and a multi-layer plain-DNN.  
    \item \textbf{ConTDM} is the proposed context-aware user preference model along with the tree index. The structural information on the tree index is incorporated by a graph convolutional layer and a parent fusion layer contained in the preference model.
\end{itemize}

We randomly sample 5,000 and 10,000 disjoint users to create testing set for Amazon Books and UserBehavior respectively. The other users in two datasets compose the training set. For each user in testing set, we take the first half of behaviors along the time line as known features and the latter half as ground truth. In negative samples generation, we deploy the same sampling strategy for all methods except Item-CF and use the same sampling ratio. 
 For fairness, both HSM and TDM use the same user preference prediction model, which contains an attention layer before a three-layer plain-DNN. In ConTDM, the parameter size of the fusion unit is taken as closely as the attention layer in TDM and the plain-DNN layers are the same as other baseline methods.  Note that we do not apply attention module to YouTube product-DNN because pairwise attention is not applicable in industrial scenario for user preference models with the inner-product form to achieve acceleration in retrieval. Besides, the same tree index learnt by the joint learning framework \cite{zhu2019joint} with a plain-DNN user preference prediction model is shared by HSM, TDM and ConTDM to make fair comparison. 

\subsection{Comparison results}

\begin{table*}[!htbp]
    \caption{Comparison results of different methods in Amazon Books and UserBehavior.} 
    \label{table:ComparisonResults}
    \centering
    \begin{tabular}{ccccccc}
        \toprule
        \multirow{2}{60pt}{\centering Method} & \multicolumn{3}{c}{Amazon Books} & \multicolumn{3}{c}{UserBehavior}\\
        \cmidrule(lr){2-4}\cmidrule(lr){5-7}
                                  & Precision       & Recall           & F-Measure       & Precision       & Recall           & F-Measure   \\
         \midrule
          Item-CF                  &     0.52\%      &      8.18\%      &   0.92\%     &   1.56\%       &    6.75\%     &      2.30\%     \\
         YouTube product-DNN      &     0.53\%      &      8.26\%      &   0.93\%     &   2.25\%       &    10.15\%     &      3.36\%     \\
         HSM                      &     0.57\%      &      8.68\%      &   1.00\%                   &   2.22\%       &   10.42\%      & 	  3.34\%     \\
         TDM                      &     0.83\%      &      13.56\%     &   1.49\%        &   3.26\%       &   15.50\%        &    4.92\%	 	 \\
         \cdashlinelr{1-7}
         ConTDM                    &    \textbf{0.87\%}      &      \textbf{14.42\%}     &   \textbf{1.55\%}        &   \textbf{3.65\%}        &   \textbf{17.12\%}        &   \textbf{5.49\%}     \\
        \bottomrule
    \end{tabular}

\end{table*}

 The quantitative results of all methods in two datasets under settings above is shown in Table \ref{table:ComparisonResults}.

 Firstly, compared with the traditional Item-CF and kNN-based retrieval models, TDM significantly improves the recommendation accuracy in all metrics.
 This result clearly shows the superiority of tree-based methods by removing the restrictions on the form of user preference modeling and enabling the use of more effective models.
 Actually, together with a carefully designed joint learning framework\cite{zhu2019joint}, TDM could outperform the brutal-force traverse of the whole corpus with the same preference model trained on raw training data only. Besides, a direct application of hierarchical softmax model to recommendation problem does not show much improvements on Item-CF and knn based method, which is consistent with the conclusion in \cite{zhu2018learning}.

 Secondly, ConTDM still outperforms the strongest baseline TDM with a 6.3\% and 10.5\% recall lift in Amazon Books and UserBehavior respectively. The comparison result shows the effectiveness of importing contexts contained on the tree index to user preference modeling. 
 Notice that the improvements are mostly achieved by the proposed context-aware user preference prediction model with a common tree index shared between different tree-based methods.
 A fine-tuned tree index and more effective training framework such as \cite{zhu2019joint,zhuo2020learning} can be naturally applied in ConTDM to achieve better overall performance.


\subsection{Ablation analysis} \label{section:Ablation}

 In ConTDM, a graph convolutional layer and a parent fusion layer are designed in the user preference prediction model to utilize both horizontal and vertical contexts on $\mathcal{T}$ respectively. 
 In this subsection, we make an ablation analysis about the effectiveness of these two layers.
 All settings are kept unchanged as the last subsection other than removing one of these two layers from $\mathcal{M}$. 
 Experimental results are shown in Table \ref{table:AblationResultsGEMTL} with TDM as the baseline. 

\textbf{\textit{Graph Convolutional Layer}}. In this case, we remove the parent fusion layer in ConTDM and keep other parts unchanged. We denote this model as \textbf{ConTDM-GC}. From Table \ref{table:AblationResultsGEMTL}, ConTDM-GC lifts the recall with the relative percentage of 6.3\% and 5.9\% in Amazon Books and UserBehavior respectively.
This result confirms the  advantage of utilizing the co-occurrence among nodes from the same level in the tree index.

\textbf{\textit{Parent Fusion Layer}}. In retrieval, the top-down path from the root to the leaf layer forms the decision chain of the user preference model, which naturally indicates the user interests granularity evolves from coarse to fine. From a probabilistic perspective, the top-down beam search process can be regarded as a sequence generation process. 
 We remove the graph convolutional layer from ConTDM and denote the preference model with the parent fusion layer as \textbf{ConTDM-PF}. Results in Table \ref{table:AblationResultsGEMTL} shows the effectiveness of the parent fusion layer. In UserBehavior, ConTDM-PF gains 6.6\% recall lift, which beats the ConTDM-GC with 5.9\%. Nevertheless, in Amazon Books, the recall yields by ConTDM-PF and TDM are roughly the same.
 We attribute this result to the impact of dataset. Since most items in this dataset are books and the coarse description of user preference on higher levels of $\mathcal{T}$ does not benefit much to prediction on child nodes.

 \begin{table}[!htbp]
	\caption{Ablation results for Graph Convolutional Layer and Parent Fusion Layer in Amazon Books and UserBehavior.} 
	\label{table:AblationResultsGEMTL}
\begin{tabular}{c|c|ccc}
\hlineB{2}
\multirow{2}{*}{Dataset}      & \multicolumn{1}{c|}{\multirow{2}{*}{Method}} & \multicolumn{3}{c}{Metric@200}  \\ \cline{3-5} 
                              & \multicolumn{1}{c|}{}                        & Precision & Recall  & F-measure \\ \hline
\multirow{4}{*}{Amazon Books} & TDM                                          & 0.83\%    & 13.56\% & 1.49\%    \\ 
                              & ConTDM-GC                                    & 0.88\%    & 14.41\% & 1.57\%    \\
                              & ConTDM-PF                                   & 0.82\%    & 13.55\% & 1.47\%    \\
                              & ConTDM                                       & \textbf{0.87\%}    & \textbf{14.42\%} & \textbf{1.55\%}    \\ \hline
\multirow{4}{*}{UserBehavior} & TDM                                          &           3.26\%	  & 15.50\% & 4.92\%     \\ 
                              & ConTDM-GC                                    &           3.50\% & 16.41\% & 5.26\%          \\
                              & ConTDM-PF                                   &           3.53\% & 16.53\% & 5.31\%          \\
                              & ConTDM                                       &           \textbf{3.65\%} & \textbf{17.12\%} & \textbf{5.49\%}           \\ \hlineB{2}
\end{tabular}
\end{table}
 
\subsection{Online Results}

 ConTDM has been applied in the display advertising scenario, i.e. Guess What You Like column of Taobao App, with full online traffic at the stage of candidate generation.
 In Taobao's advertising systems, advertisers bid on the reveals that show items to users. 
 When a user opens the Taobao App, the advertising engine should choose a few proper ads from the large scale corpus of ads to be revealed. 
 Generally, the whole process in practice can be devided into three subsequent stages: candidate generation, ranking and strategy. 
 After each stage, the candidate set is reduced gradually from the whole corpus containing millions of items to few ads. 
 Besides, the target considered in each stage also varies to meet different business goals. 

 To measure the performance, we conduct online A/B comparison by replacing ConTDM with the strongest baseline  method TDM. Each comparison experiment has 2\% of all online traffic.
 We use click-through rate (CTR) and revenue per mille (RPM) that are the key performance indicators for online display advertising as metrics. The definitions of these two metrics are as follows:

\begin{align*}
    \text{CTR} = \frac{\text{\# of clicks}}{\text{\# of impressions}},~\text{RPM} = \frac{\text{Ad revenue}}{\text{\# of impressions}} * 1000.
\end{align*}
 Besides, the diversity defined by the size of different categories in the candidate set is also considered.  

\begin{table}[!htb]
    \caption{Online results in \emph{Guess What You Like} column of Taobao App Homepage.}
   \label{table:OnlineResults}
   \centering
   \begin{tabular}{rcccc}
       \toprule
       & Metric & CTR & RPM & Diversity \\
       \midrule
       & ConTDM & +3.8\% & +4.0\% & +14.0\% \\
       \bottomrule
   \end{tabular}
\end{table}

 Table \ref{table:OnlineResults} reveals the lift on all online metrics. 3.8\% growth on CTR exhibits that more precise items have been recommended. RPM with the increase of 4.0\% proves ConTDM can bring more income for Taobao advertising platform. Thanks to the horizontal contexts brought by the hierarchical graph, the diversity of candidate set is also significantly improved by 14.0\%. It means more potential improvements can be made by subsequent stages since their corpus is largely enriched.  
 Notice that there are sevaral different methods working simultaneously online at the candidate generation stage and ConTDM is only one of them. Besides, the cancandidate set returned in the  candidate generation stage will be reranked by the subsequent stages to pick the few final ads. Therefore, the improvements achieved by ConTDM is very significant.
 Besides, as discussed in Section \ref{sec:gclayer}, graph embeddings of ConTDM are aggregated offline and all improvements compared with TDM are achieved without increase of the stress of the online engine. 

%% file: section6.tex
\section{Conclusion}\label{section:Conclusion}

 In this paper, we study the effect of the tree index in user preference modeling and propose a context-aware user preference prediction model, which can be conveniently applied in general tree-based methods for recommender systems.
 Both horizontal and vertical contexts on $\mathcal{T}$ are utilized through a novel graph convolutional layer and a parent fusion layer. 
 Both online and offline results show the significant improvements brought by ConTDM.

 In ConTDM, we mainly focus on how to improve the recommendation accuracy of user preference model in tree-based methods.  
 Actually, the quality of the tree index also matters \cite{zhu2019joint}. 
 Intuitively, we can naturally extend the binary tree index used in ConTDM to a multi-path tree index with the constructed hierarchical graph  
 and its capacity is much increased.
 We will further study how to make effective and efficient training and retrieval on the more challenging multi-path tree index in our future work. 